# Psychological Factors Influencing University Students' Trust in AI-Based Learning Assistants

**Ezgi Dagtekin[1], Ercan Erkalkan[2]**

[1]The MacDuffie School, Massachusetts, USA
[2]Department of Electronics and Automation, Marmara University, Istanbul, Türkiye
Corresponding author: Ercan Erkalkan (e-mail: ercan.erkalkan@marmara.edu.tr)

**Abstract**

Artificial intelligence (AI)-based learning assistants and chatbots are increasingly integrated into higher education. While these tools are often evaluated in terms of technical performance, their successful and ethical use also depends on psychological factors such as trust, perceived risk, technology anxiety, and students' general attitudes toward AI. This paper adopts a psychology-oriented perspective to examine how university students form trust in AI-based learning assistants. Drawing on recent literature in mental health, human AI interaction, and trust in automation, we propose a conceptual framework that organizes psychological predictors of trust into four groups: cognitive appraisals (perceived competence, reliability, and usefulness), affective reactions (anxiety, fear, comfort), social-relational factors (anthropomorphism, perceived empathy, and autonomy support), and contextual moderators (privacy, transparency, and ethical norms). A narrative review approach is used to synthesize empirical findings on AI tools in education and mental health support, with a specific focus on college and university populations. Based on this synthesis, we derive research questions and hypotheses for future survey and experimental studies with students. The paper highlights that trust in AI is not a purely technical outcome but a psychological process shaped by individual differences and learning environments. Finally, practical implications are discussed for instructors, university administrators, and designers of educational AI systems, including guidelines to foster appropriate reliance rather than blind trust or total rejection.

## 1. INTRODUCTION

Artificial intelligence (AI)-based systems such as chatbots, recommendation engines, and conversational agents are now widely used in education and mental health support for university students. Recent systematic reviews show that AI-based conversational agents can reduce symptoms of depression, anxiety, and psychological distress when used as supportive tools, especially among young adults and students [1–3]. However, these reviews also emphasize that the effectiveness and safety of such tools depend on how users perceive and relate to AI, not only on algorithmic accuracy [4].
From a psychological perspective, AI systems are not neutral technologies. Students interpret AI tools through existing beliefs, emotions, and social experiences. Concepts such as trust, perceived fairness, technology anxiety, and sense of control play a decisive role in whether students choose to rely on AI recommendations or actively reject them. Inadequate trust may lead to underuse of potentially helpful tools, whereas inappropriate trust can create overreliance and neglect of critical thinking in learning or health-related decisions [5].

Despite the rapid diffusion of AI-based learning assistants for example, AI-powered homework helpers, writing support tools, or course-specific chatbots many studies still emphasize technical performance rather than psychological processes. There is a need for psychology-oriented frameworks that explain how and why students trust or mistrust AI and which psychological factors predict appropriate reliance [1, 6, 7].

The main purpose of this paper is to develop a psychology centered conceptual framework for understanding university students' trust in AI-based learning assistants. The contribution of the paper is threefold. First, it synthesizes empirical findings on AI tools in education and mental health focusing on college and university populations. Second, it identifies key psychological constructs that influence trust and reliance on AI. Third, it proposes research questions and hypotheses that can guide future empirical work in the "Social, Humanities, and Administrative Sciences" domain.

The remainder of the paper is organized as follows. Section 2 presents the material and method, describing the narrative literature review strategy and the psychological constructs considered. Section 3 introduces the proposed framework and summarizes the main themes identified in the literature. Section 4 concludes with implications, limitations, and suggestions for future research.

## 2. MATERIAL AND METHOD

This study follows a narrative literature review approach rather than a systematic meta-analysis. The aim is not to estimate pooled effect sizes, but to organize and interpret recent psychological evidence regarding trust in AI among students and young adults. Narrative reviews are suitable when the goal is to integrate findings from diverse study designs and to propose conceptual frameworks rather than to test a single, predefined model statistically.

To identify relevant research, major scholarly databases and publishers (such as IEEE Xplore, ScienceDirect, SpringerLink, PubMed, and Frontiers) were consulted using combinations of the following keywords in English: artificial intelligence, chatbot, conversational agent, trust, psychology, university students, mental health, technology acceptance, and human-AI interaction. The focus was on studies and reviews published between 2018 and 2025 that satisfied at least one of the following criteria: (i) The sample consisted primarily of university or college students, or of young adults in an educational context; (ii) the study examined psychological outcomes of AI use (e.g., attitudes, trust, anxiety, well-being, or motivation); or (iii) the study dealt with AI-based conversational agents or learning support tools relevant to students [8, 9].

Recent reviews on AI-driven conversational agents in mental health and digital well-being were used as anchor sources for identifying further empirical work [8, 9]. In addition, reviews on trust in AI and trustworthy AI from a psychological viewpoint were included to understand how trust has been conceptualized and measured [4–6]. As this paper is based on a literature review and does not involve new data collection from human participants, formal ethics committee approval was not required.

### 2.1. Psychological Constructs Considered

Based on the selected literature, four groups of psychological constructs were considered when organizing the proposed framework. The first group consists of cognitive appraisals, including perceived competence, accuracy, reliability, usefulness, and fairness of AI systems. The second group covers affective reactions such as technology-related anxiety and fear, curiosity, comfort, and perceived emotional safety when interacting with AI. The third group relates to social-relational factors, such as anthropomorphism (seeing AI as humanlike), perceived empathy, interpersonal warmth, and perceived autonomy support in learning. Finally, contextual moderators include privacy concerns, perceived transparency and explainability, norms in the educational environment, and ethical regulations. These constructs are frequently mentioned in psychological research on human-AI trust and are grounded in social-cognitive models of trust and technology acceptance [4, 5].

Table 1. Summary of inclusion and exclusion criteria and study designs used in the narrative review

| Criterion Category | Description | Representative Examples of Studies or Typical Applications |
|---|---|---|
| Inclusion criteria | Peer-reviewed publications between 2018 and 2025 focusing on university students or young adults and reporting psychological constructs related to AI-based systems | Narrative and empirical studies on trust, attitudes, perceived risk, technology anxiety, motivation, or well-being in relation to AI-based learning assistants |
| Exclusion criteria | Publications without psychological measures, studies outside higher education, or articles lacking sufficient methodological information | Technical system descriptions without psychological outcomes; studies based on from kindergarten to 12th grade (K–12) samples; reports not published in peer-reviewed venues |
| Study designs | Experimental interventions, survey-based quantitative studies, and qualitative or mixed-method investigations | Controlled experiments manipulating properties of AI-based assistants; large-scale online surveys in higher-education settings; interview-based or focus-group studies on students' experiences with AI-supported learning |

### 2.2. Inclusion and Exclusion Criteria and Study Designs

The narrative review focuses on peer-reviewed publications released between 2023 and 2025. The selection strategy follows explicit inclusion and exclusion criteria in order to ensure conceptual relevance to university students and AI-based learning assistants.

- Inclusion criteria: Empirical or review article published in a peer-reviewed journal, conference proceedings, or edited volume between 2023 and 2025.
- Inclusion criteria: sample composed primarily of university or college students or young adults in an educational context.
- Inclusion criteria: Focus on psychological constructs related to AI-based systems, such as trust, attitudes, perceived risk, technology anxiety, motivation, or well-being.
- Inclusion criteria: Use of AI-based conversational agents, learning assistants, recommendation tools, or comparable educational AI systems.
- Exclusion criteria: Publications limited to technical performance, algorithm design, or system architecture without psychological measures.
- Exclusion criteria: Studies conducted exclusively with clinical populations, K–12 students, or general consumers outside higher education.
- Exclusion criteria: Articles lacking sufficient methodological information or not available in English.

The final body of evidence covers three main study designs: experimental interventions that expose students to AI-based tools under controlled conditions, survey-based quantitative studies that measure psychological constructs in naturalistic educational settings, and qualitative or mixed-method investigations that explore subjective experiences with AI-supported learning.

## 3. FINDINGS AND PROPOSED FRAMEWORK

In the context of this narrative review, "results" refer to the main conceptual themes and relationships identified in the literature rather than statistical findings from a single empirical dataset. The proposed framework suggests that trust in AI-based learning assistants among university students emerges from the interaction of cognitive appraisals, affective reactions, social relational factors, and contextual moderators.

Figure 1 organizes the conceptual framework around trust in AI-based learning assistants. The central node represents students' trust and appropriate reliance on AI, surrounded by four clusters of predictors: cognitive appraisals, affective reactions, social-relational factors, and contextual moderators. Directed arrows indicate expected positive or negative associations between each cluster and trust, as well as moderating influences of contextual factors on the links between cognitive, affective, and social-relational variables and trust.

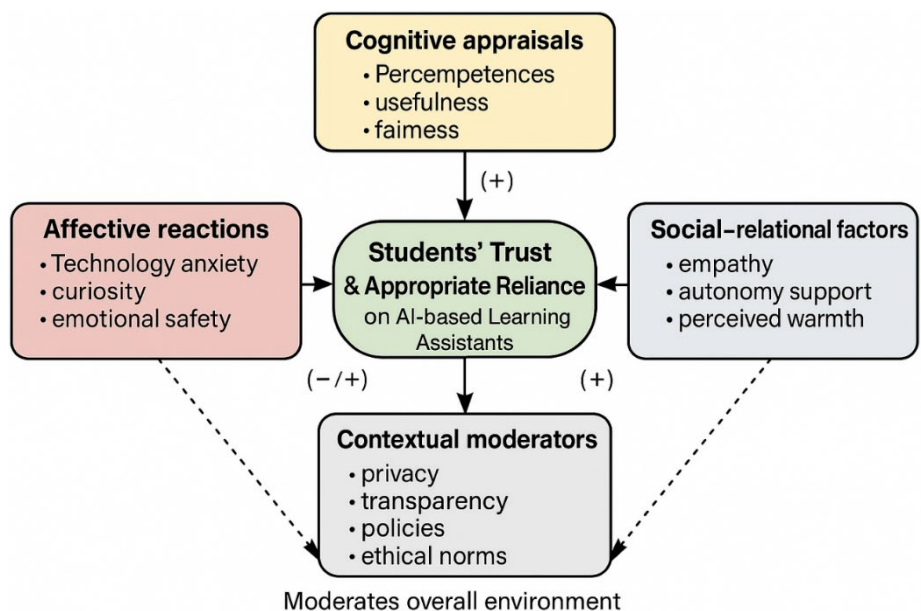

**Figure 1.** Psychology-oriented framework of university students' trust in AI-based learning assistants

Table 2 provides a concise overview of how key psychological construct clusters have been operationalized across different study designs and summarizes the dominant empirical associations with trust in AI-based learning assistants.

**Table 2.** Key psychological constructs, typical study designs, and dominant empirical patterns related to trust in AI-based learning assistants

| Construct Cluster | Representative Variables | Typical Study Designs | Dominant Empirical Patterns |
|---|---|---|---|
| Cognitive appraisals | Perceived competence, accuracy, usefulness, fairness | Experimental interventions; survey-based quantitative studies | Higher cognitive appraisals associated with increased trust, stronger continuance intention, and higher perceived learning support |
| Affective reactions | Technology anxiety, fear, curiosity, emotional safety | Survey-based cross-sectional and longitudinal designs; experimental studies | Higher anxiety associated with lower trust and reduced usage; higher emotional safety associated with greater engagement and willingness to experiment with AI tools |
| Social-relational factors | Anthropomorphism, perceived empathy, autonomy support, interpersonal warmth | Survey-based studies; laboratory experiments; qualitative interviews | Supportive social cues associated with stronger perceived empathy and more appropriate reliance, particularly when autonomy support remains salient |
| Contextual moderators | Privacy concerns, transparency, institutional policies, ethical norms | Survey-based studies; policy analyses; mixed-method research | Clear policies and transparent communication associated with higher trust and a weaker negative impact of privacy concerns on usage intentions |

### 3.1. Cognitive Appraisals: Perceived Competence, Reliability, and Usefulness

A consistent finding in the literature is that users are more likely to trust and rely on AI tools when they perceive them as competent and reliable. In educational settings, this means that students must believe the AI assistant provides correct, up-to-date, and relevant information. Studies on conversational agents for mental health similarly show that perceived accuracy and helpfulness are strongly associated with continued use [1–3, 9]. Perceived usefulness in supporting learning outcomes such as clearer explanations, improved grades, or time savings also plays a central role. When students experience that an AI assistant helps them clarify complex concepts or structure their study routines, trust tends to increase. Conversely, if AI suggestions frequently conflict with course materials or instructor feedback, students may downgrade the system's reliability and reduce their reliance.

### 3.2. Affective Reactions: Anxiety, Fear, and Emotional Safety

Beyond rational evaluations, affective reactions strongly shape trust in AI. Psychological research indicates that technology anxiety and fear of making serious mistakes with the help of AI can reduce willingness to engage with AI tools, even when their technical performance is objectively high [7]. For some students, AI-based assistants may reduce anxiety by providing non-judgmental, always-available support. For others, especially those with concerns about surveillance or job displacement, AI can evoke worry, anger, or feelings of dehumanization [1, 2]. In mental health contexts, reviews highlight both the potential of chatbots to alleviate symptoms and the risk that emotionally vulnerable users might receive responses that feel invalidating or unsafe [1–3, 8, 9]. From a psychological standpoint, emotional safety the feeling that one can interact with an AI assistant without being ridiculed, monitored excessively, or harmed is a critical precursor to trust.

### 3.3. Social-Relational Factors: Anthropomorphism, Empathy, and Autonomy Support

Many students implicitly treat AI systems as social partners, especially when they are designed with humanlike language, names, or avatars. Anthropomorphism can increase perceived empathy and warmth, which may enhance trust and engagement. However, if the system's behaviour does not match these expectations for example, by giving generic or insensitive responses users may feel disappointed or even betrayed [4, 5, 8]. For educational AI assistants, perceived autonomy support helping students make their own decisions instead of prescribing answers appears to be particularly important. When AI tools explain options, encourage reflection, and respect students' choices, they reinforce a sense of agency and intrinsic motivation, consistent with self-determination theory.

### 3.4. Contextual Moderators: Privacy, Transparency, and Ethical Norms

Trust in AI is also shaped by the broader context in which students encounter these systems. Transparency about data usage, storage, and model limitations is known to influence perceived trustworthiness [4, 5]. In university settings, students may have additional concerns: whether their interactions with AI-based learning assistants are monitored by instructors or administrators; how AI-generated content is evaluated in terms of academic integrity and plagiarism; and whether the institution provides guidelines or training on responsible AI use. The presence of clear ethical norms and institutional policies may reduce uncertainty and support appropriate reliance that is, using AI when it is helpful but still engaging in critical thinking and seeking human support when needed.

### 3.5. Proposed Research Questions and Hypotheses

Based on the reviewed literature and the proposed framework, several research questions (RQs) and illustrative hypotheses (Hs) can guide future empirical work with university students:

- RQ1: How do cognitive appraisals such as perceived competence, reliability, and usefulness predict students' trust in AI-based learning assistants?
  H1: Higher perceived competence and usefulness will be positively associated with trust and intended continued use.
- RQ2: What is the role of affective reactions such as technology anxiety, fear, and emotional safety in shaping trust?
  H2: Technology anxiety will be negatively associated with trust, controlling for perceived competence.
- RQ3: How do social relational factors such as anthropomorphism, perceived empathy, and autonomy support—influence appropriate reliance on AI?
  H3: Perceived empathy and autonomy support will predict appropriate reliance, defined as using AI as a supportive tool without fully delegating decisions.
- RQ4: How do contextual moderators such as privacy concerns, perceived transparency, and institutional policies influence the relationship between psychological factors and trust?
  H4: High perceived transparency and strong institutional guidelines will weaken the negative effect of privacy concerns on trust.

Each research question in the proposed set corresponds to a specific group of links in the conceptual framework. RQ1 addresses pathways from cognitive appraisals to trust in AI-based learning assistants. RQ2 focuses on affective reactions, including technology anxiety and emotional safety. RQ3 targets social-relational factors and appropriate reliance. RQ4 examines the moderating influence of contextual factors such as privacy concerns, transparency, and institutional policies on the associations between cognitive, affective, and social-relational variables and trust.

These questions can be investigated through cross-sectional surveys, longitudinal designs, or controlled experiments in which students interact with specific AI-based learning tools in real or simulated course settings.

## 4. CONCLUSION

This paper has developed a psychology-oriented framework for understanding university students' trust in AI-based learning assistants. Rather than focusing solely on algorithmic performance, the framework emphasizes that trust is a multidimensional psychological process shaped by cognitive appraisals, affective reactions, social relational factors, and contextual moderators. The narrative review of recent research in mental health, human AI interaction, and trust demonstrates that AI tools can support students' learning and well-being, but only when psychological needs such as competence, autonomy, and emotional safety are respected [1–6, 8, 9].

For practitioners, including designers and educators, the framework suggests several practical implications. First, they should provide transparent information about what AI tools can and cannot do. Second, interactions should be designed to support autonomy and critical thinking rather than passive dependence. Third, students' emotional concerns such as anxiety and fear about AI should be openly discussed and addressed. Finally, universities should establish clear institutional policies on privacy, academic integrity, and responsible AI use so that students can rely on AI tools appropriately.

## 4.1. Implications for Higher Education in Türkiye

Higher-education institutions in Türkiye currently experience rapid diffusion of AI-based tools in both formal and informal learning environments. University students frequently engage with global large language models and locally developed educational platforms without systematic guidance regarding trustworthy and responsible use. The proposed framework highlights several priorities for institutional policy and course design in this context.

First, measurement of psychological constructs such as trust in AI, technology anxiety, and perceived autonomy support requires culturally adapted scales. Existing instruments from the technology acceptance and trust in automation literature can be translated and validated for Turkish university students through standard psychometric procedures, including exploratory and confirmatory factor analysis. Rigorous adaptation will support longitudinal monitoring of students' attitudes and perceived risks.

Second, curriculum design in Türkiye can incorporate explicit discussions of academic integrity, plagiarism, and appropriate reliance on AI-based learning assistants. Clear course-level rules, sample use cases, and reflective assignments can align students' expectations with institutional norms. Courses in fields such as engineering, social sciences, and teacher education can embed short modules on critical evaluation of AI-generated content.

Third, university-level governance can develop transparent policies regarding data protection, logging of student–AI interactions, and support services. Collaboration between information technology units, ethics committees, and counseling centers can ensure that AI-based tools complement rather than replace human support in mental health and academic advising. Context-sensitive policy design in Türkiye will enable university students to benefit from AI-based learning assistants while maintaining psychological safety and trust.

## Acknowledgments

The author would like to thank colleagues and students at Marmara University for their feedback on early versions of this work.